\documentclass{article}

\usepackage{graphicx} 
\usepackage{url}      
\usepackage{multirow}
\usepackage{listings}
\usepackage{algorithmic}
\usepackage{algorithm2e}

\begin{document}

\title{Virtual Screening on FPGA: Performance and Energy versus Effort}

\author{
     Tom Vander Aa,  
     Tom Haber, 
     Thomas J. Ashby, \\
     Roel Wuyts, and 
     Wilfried Verachtert \\
     ExaScience Life Lab at imec, Leuven, Belgium \\
     Email: tom.vanderaa@imec.be
}

\maketitle

\begin{abstract}
    With their wide-spread availability, FPGA-based accelerators cards have
    become an alternative to GPUs and CPUs to accelerate computing in
    applications with certain requirements (like energy efficiency) or
    properties (like fixed-point computations). 
    In this paper we show results and experiences from mapping an industrial
    application used for drug discovery on several types of accelerators. We especially 
    highlight the effort versus benefit of FPGAs compared to CPUs and GPUs in
    terms of performance and energy efficiency.
    For this application, even with extensive use of FPGA-specific features, and
    performing different optimizations, results on GPUs are still better, both in 
    terms of energy and performance.
\end{abstract}

\section{Introduction}
  
Today FPGAs are proposed as  an alternative to GPUs or CPUs to speed-up
computation. They are available in the datacenter \cite{aws_f1}, and
supercomputers are being build with FPGAs as their main compute elements
\cite{euroexa}



With their massive parallelism, high-Bandwidth memory \cite{FPGA_hbm} and low
energy consumption \cite{float_kernel}, they have clear benefits in certain
situations to CPUs and GPUs. Also, in recent years, FPGA floating point
performance has improved \cite{FPGA_float} and fixed point performance is even
better.

While in the past you could only program FPGAs in low-level hardware description
languages like Verilog or VHDL, these days using high-level programming in C or C++
and OpenCL is the norm. Indeed, high-level synthesis (HLS) tools have improved
dramatically over the last years \cite{hls_survey}, and the use of so-called pragma's
allows a cleaner separation of pure functionality and FPGA-specific
optimizations. Furthermore, both Intel and Xilinx have made their tools
available without requiring a license, and have invested heavily in libraries to
accelerate common compute tasks on FPGA (Intel oneAPI \cite{oneapi} and Xilinx
Vitis \cite{vitis}).

It has never been easier to get your first kernel running on an FPGA accelerator
card.  But do you get good performance and low energy consumption for
compute-intensive applications?  And how much time do you need to invest to
optimize your code?  How do these metrics compare to GPU and CPU
implementations? In this paper we try to answer these questions studying a
Virtual Molecule Screening (VMS) application.

This paper is organized as follows. In Section~\ref{sec:related} we
look at related  work in FPGA mapping and performance optimizations.
Section~\ref{sec:vms} introduces the VMS applications, which is
optimized in Section~\ref{sec:steps}. In Section~\ref{sec:results}, 
we look at performance numbers on FPGA and compare those to equivalent
GPU and CPU versions. Sections~\ref{sec:conclusions} are the conclusions.


\section{Related Work}
\label{sec:related}

High-Level Synthesis (HLS) has improved so much over the last year that it has
become the de facto standard for FPGA acceleration \cite{hls_survey}. HLS avoids
the writing and verification of the functionality in low-level Hardware
Description Language (HDL) like VHDL or Verilog. Yet it is known that such
low-level HDL code can be fine-tuned for the FPGA resulting in much higher clock
frequency, much better resource utilization and thus much better performance
\cite{supertile}.

Alternatively, in the domain of machine learning, many frameworks exits to
automatically generate HLS code, with optimizations for FPGA included
\cite{Bouganis}. Such frameworks are typically tuned to a specific subdomain,
for example convolutional neural networks \cite{fpgaConvNet}.

Older publications evaluating the mapping and optimization of compute-intensive
kernels on FPGA show a clear energy advantage compared to GPU and CPU
implementations \cite{float_kernel}. Recently with the advent of larger GPUs and
CPUs,  FPGA only seem to excel in latency and seem outperformed by GPUs both in
energy efficiency and performance \cite{fpga_trends}.  Loop transformations have
always played a crucial role in optimizing HLS code \cite{hls_loop_trafos}.


\section{Virtual Molecule Screening}
\label{sec:vms}

In this section we explain some key characteristics of the Virtual Molecule Screening
application.
   
\subsection{Chemogenomics and Matrix Factorization}

In chemogenomics the key problem is the identification of candidate molecules
that affect proteins associated with diseases. If this is done using machine
learning techniques, the process is called compound-activity prediction \cite{excape_db}.
Bayesian Matrix Factorization (BMF \cite{BPMF}) is a technique borrowed from recommender
systems that has been successfully used for compound activity prediction.
Thanks to the Bayesian approach, BPMF has been proven to be more robust to
data-overfitting and Gibbs sampling makes the models feasible to compute.
Another benefit, compared to for example deep learning, is that predictions from
BMF models include confidence estimation, that can be used as a quality metric.

\subsection{Virtual Screening}

\begin{figure} \centering \includegraphics[width=\columnwidth]{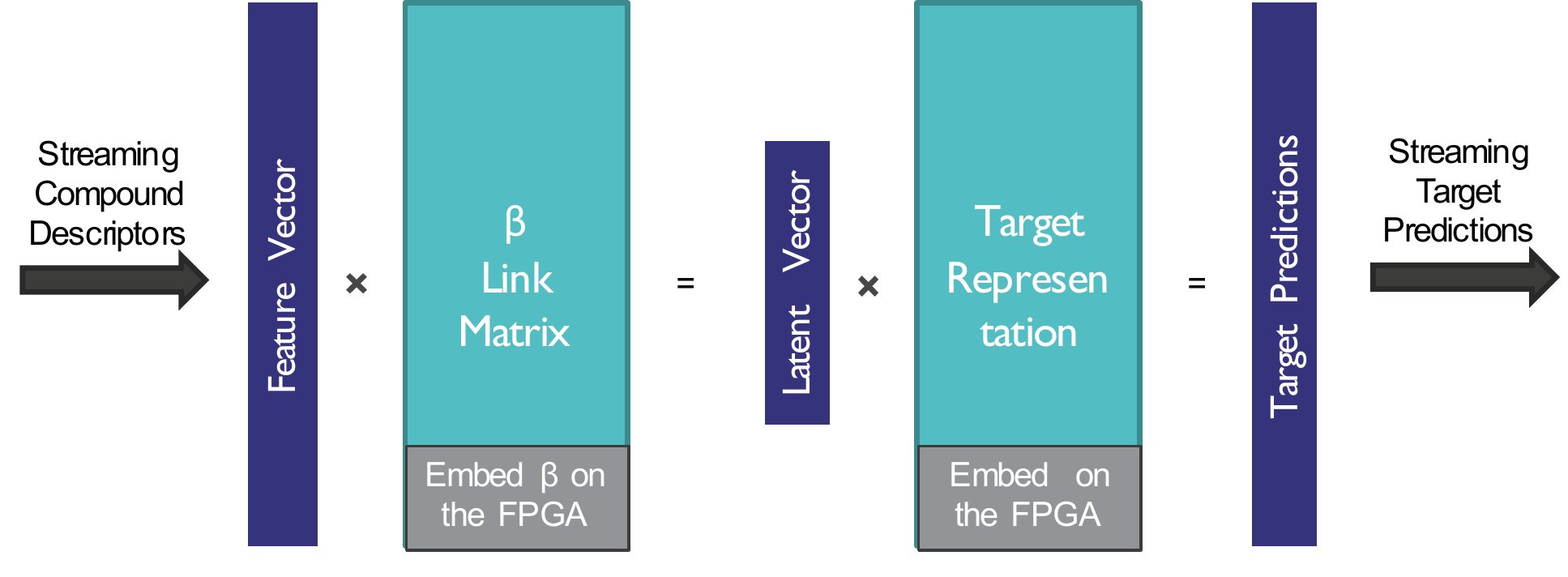}
\caption{Virtual Molecule Screening (VMS) Pipeline}
\label{fig:vms} \end{figure}

Once the models have been built, drug companies want to use them to evaluate
millions of molecules in so-called virtual screens.  Figure~\ref{fig:vms} shows
a simplified view on the prediction flow for activity prediction. A molecule
represented by its chemical fingerprint is fed in on the left. From the
fingerprint an internal latent representation is computed, and this latent
representation is used to make predictions on one or more protein targets. The
amount of computation depends on the number of molecules, the number of Gibbs
samples, the size of the latent representation and the number of protein
targets. Since many compounds need to be tested, virtual screening requires much
computational power, justifying the need for acceleration.

\subsection{Virtual Screening Kernel}

The main part of the VMS application is its kernel. More than 90\% of the
runtime is spent in this part of the application. To prepare for acceleration we
have rewritten the kernel in HLS-ready C++. We have wrapped it in an OpenCL
interface, to make it suitable for offloading.

\section{Accelerating VMS using FPGAs}
\label{sec:steps}

As already discussed in Section~\ref{sec:related}, there are clear benefits and
challenges related to FPGAs.  In this section we describe the different steps to
exploit, avoid or remedy these benefits and challenges, by optimizing and
rewriting our VMS application. In the next section we refer back to these steps
when we evaluate performance, energy efficiency and optimization effort.

The steps are present here from most generic optimization, towards most targeted
towards the actual FPGA device and the prediction model used. This order happens 
to coincide with going from the inner parts of the kernel, the lowest level of
optimizations (the operations), over the loops, and functions of the kernel,
outwards to the kernel as a whole and even beyond the kernel to the host
interface.

\subsection{Code Complexity}

Mapping code efficiently on FPGA is a time-consuming task, even with help of
high-level synthesis tools. Luckily for us, the VMS application code base is small
and simple.

The structure of the kernel is a set of nested for loops. We have applied loop
blocking \cite{blocking}.  By carefully choosing the block-size, we can
fine-tune the kernel parallelism to match the available resources of the FPGA.

\subsection{Bit-Width Reduction}

FPGAs deal much better with reduced bit-width fixed point numbers than with
double or single floating point numbers \cite{FPGA_float}.  We were able to
reduce the bit width of the input and output streams, and of the model itself
from 64 bit floating point to 16bit and 8bit fixed point, without a significant
loss in prediction accuracy. We did this using profiling and automatic fixed
point refinement \cite{fixed_point_refinement}.

\subsection{Parallelism}

Compared to CPUs, FPGAs have a clock-speed that is 10x lower than CPUs. They
compensate for this by providing many more parallel resources (DSP blocks,
memory's, LUTs, ...). We were able to benefit from these extra resources by
exploiting parallelism at many levels: inside the prediction pipeline, but also
across proteins and molecules.

The use of Xilinx HLS pragmas allowed use to perform these optimizations
unobtrusively, and allowed us to explore many ways to exploit the
parallelism, at loop level (using loop unrolling, and loop pipelining), and at
the function level using dataflow parallelism with FPGA-based state machines.

Since compute now happens distributed, data needs to be distributed as well.
Data distribution mirrors the loop transformations and is implemented using the
pragma \emph{array partition}. The data distribution for the dataflow across
functions is implemented using HLS streams.

\subsection{Memory Bandwidth}

As with all accelerators, getting data in and out efficiently is key to good
performance. In this case we achieved this by streaming the fingerprints in and
predictions out linearly, and by storing the model itself in the FPGA
on-chip memory beforehand. 

Streaming was implemented by fine-tuning the AXI interface to the kernel. This
allowed us to do AXI pipelined burst accesses and use the full bitwidth (512bits)
of the memory interface.

\subsection{Kernel Dimensions}

To make most efficient use of the FPGA resources we have to make sure the kernel
matches the amount of each of them available. We can influence this by changing
the kernel dimensions, for example, by choosing how many compounds, or samples
we process in parallel. We have also optimized the number of compounds per kernel
invocation to reduce the kernel invocation overhead, and increase the efficiency
of the dataflow and loop pipelining.

\subsection{Kernel Interface}

The final step is outside the kernel itself, at the level of the OpenCL interface.
This involves instantiating multiple instances of the  kernel to make optimal use multiple 
regions on the FPGA (SLRs \cite{xilinx_slr}), each with their own DRAM memory interface.

We also make sure to overlap communication and computation at this level by using asynchronous
OpenCL call \cite{xilinx_ug1393}.

\section{Results and Discussion}
\label{sec:results}

In this section we evaluate the different optimization steps, we compare
the final results to a GPU and CPU implementation and discuss
performance versus effort for the three implementations.

\subsection{Optimization Steps}

\begin{figure}
  \centering
  \includegraphics[width=\columnwidth]{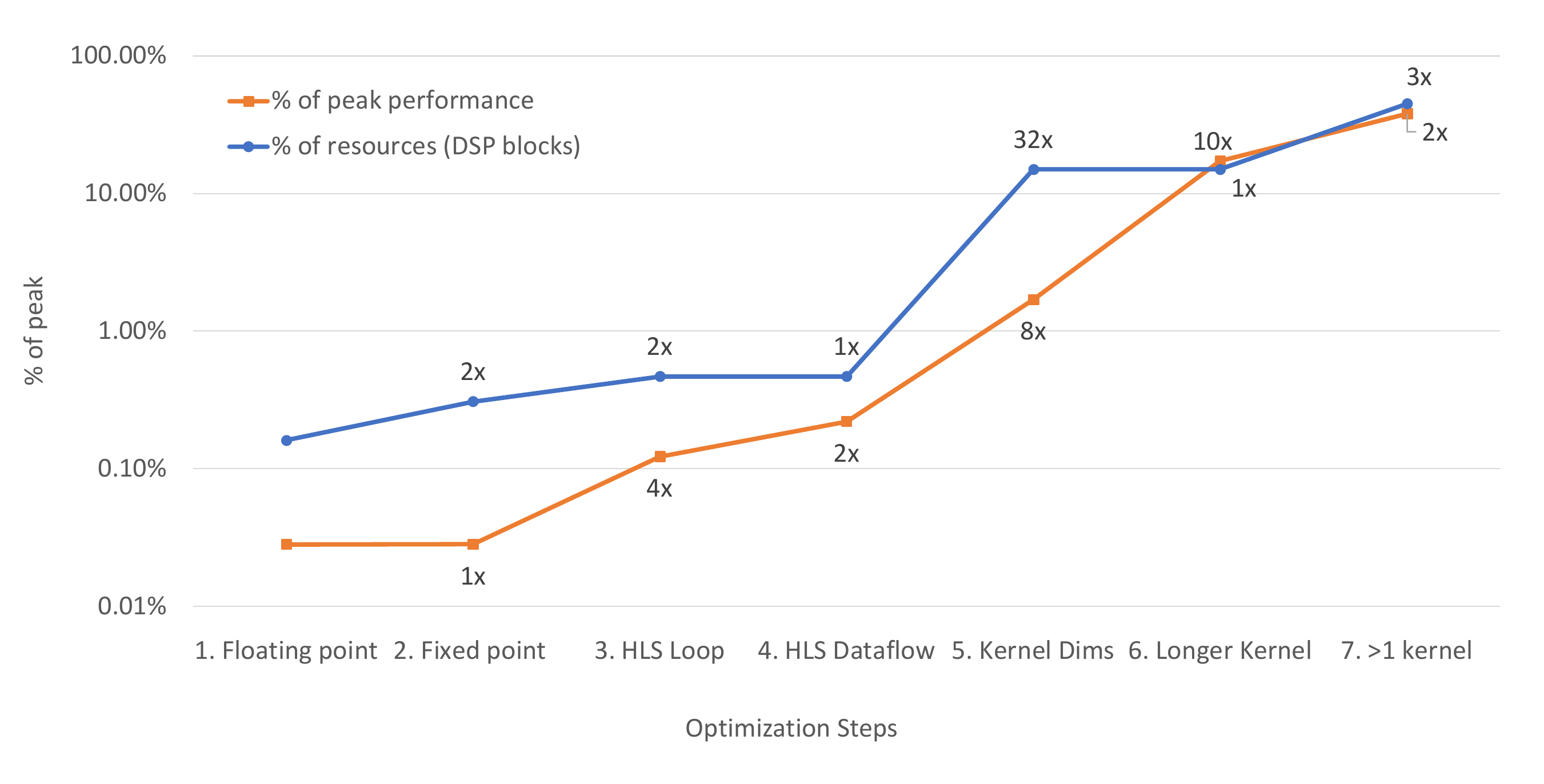}
  \caption{Effect of the Different optimization steps on performance (as a percentage of peak performance)
  and on resource usage (as a percentage of DSP blocks used)}
  \label{fig:steps} \end{figure}

We evaluated the aforementioned optimization steps on a Xilinx Alveo U200
\cite{alveo_u200} datacenter card on the VMS application.
Figure~\ref{fig:steps} shows the effect of each of the step on performance and
resource usage. Both are calculated as a percentage of peak based on the number
of available DSP slices.  Noting that the Y-axis is logarithmic, each step
increased the performance with a factor, indicated on each point in the graph.
The total performance increases with a factor $1351\times$ and the total
resource usage with $280\times$, indicating the crucial importance of optimizing
the generated C++ code.

The highest improvement comes from the step that tunes the kernel dimensions
because that step enables the full parallelism of the FPGA. While the steps
preceding this step also have a significant positive effect on performance, they
are also important enabler steps to \emph{Kernel Dimensions} step. They increase
the effect of this step.

\subsection{Comparison}

Two alternative implementations were of VMS made, one for GPU
using the ArrayFire \cite{arrayfire} library, one for CPU using
the Eigen \cite{eigen} library. We spent significant effort
to make sure these implementations perform well on their respective
platform.

Performance numbers for GPU and CPU were collected on the Juwels supercomputer 
installed at Forschungszentrum Jülich, energy numbers for GPU and CPU on the
COKA system at the University of Ferrara. FPGA and CPU energy consumption has
been monitored using hardware counters available in the system, while the GPU
power drain could be monitored using the Nvidia NVML library.

Table~\ref{tbl:cpu_gpu_fpga} shows the results. Peak performance of the three
systems was calculated using the maximum number of multiply-accumulate
operations per cycle, multiplied by the clock frequency. For the FPGA, we could
reliably reach a clock frequency of 100Mhz.

While the three systems clearly have a different peak performance in giga-flops
per second (GF/s), the three implementations reach a similar portion of
peak performance (\%peak). The FPGA implementation's percentage of peak is
higher because it exploits low-precision fixed-point calculations, which are
more efficient on FPGA. Exploiting fixed-point on GPU or CPU did not improve
performance.  

In terms of energy consumption, the GPU is clearly the winner, the CPU
performs worst and the FPGA is in between. This can be explained by look at the
peak-performance to power consumption ration. While the FPGA power consumption
is an order of magnitude lower compared to the CPU and GPU, its peak performance is
two orders of magnitude smaller compared to the GPU system. 

\begin{table}[!ht]
    \caption{Comparison of energy performance of the VMS application 
    implemented on an Nvidia A100 GPU, an Intel Skylake CPU (24 cores @ 2.7Ghz) and a
    Xilinx Alveo U200 FPGA}
    \label{tbl:cpu_gpu_fpga}
    \centering
    \begin{tabular}{l|rrr}
        ~ & CPU & GPU & FPGA \\
    \hline
        Peak Performance (GF/s) & 3072 & 19500 & 684 \\ 
        Achieved Performance (GF/s) & 402 & 3265 & 260 \\ 
        \% of Peak Performance & 13\% & 17\% & 38\% \\ 
    \hline
        Measured Power Drain (Watt) & 205 & 200 &  37 \\ 
        Energy Efficiency (GF/s/Watt) & 1.8 & 10 & 3 \\ 
    \end{tabular}
\end{table}

\subsection{Optimization Effort}

Let us conclude this section with some estimates on effort spent optimizing
the different implementations. For the CPU and GPU implementations we were
able to use optimized libraries, for the FPGA implementation we were unable to
reach good performance with the Vitis Libaries, which were available only late in
the implementation of VMS.  

Hence, we spent most effort optimizing the FPGA implementation, logging around
600 commits in version control, over three years of development. For the CPU
and GPU versions combined we registered 30 commits over a period of two years.

Additionally, long compilation times for the FPGA version (several hours per
run, unless we could rely on performance estimates after high-level synthesis), with
frequent compilation failures due to routing congesting also hampered the
optimization process.

\section{Conclusions}
\label{sec:conclusions}

Optimizing your compute-intensive application for FPGA acceleration
can improve performance several orders of magnitude, $1000\times$ for
our application. But even with FPGA-specific optimizations, like using
fixed-point, the final result is worse than a simpler GPU version, both
in terms of performance and energy consumption. While this is somewhat
expected \cite{fpga_trends}, one has to wonder if targeting FPGAs to 
accelerate compute is worth the effort with current generation hardware.

\section*{Acknowledgments}
The authors would like to acknowledge funding from the European Union's Horizon
2020 Research and Innovation programme under Grant Agreement no. 754337
(EuroEXA).  This research received funding from the Flemish regional government
(AI Research Program).  We acknowledge PRACE for awarding us access to JUWELS
at GCS@FZJ, Germany.  Many thanks to Enrico Calore for giving us
access to the COKA Cluster at INFN and the University of Ferrara.

\bibliographystyle{abbrv}
\bibliography{vms}

\end{document}